\documentclass[12pt,english]{article}
\usepackage[T1]{fontenc}
\usepackage[latin9]{inputenc}
\usepackage{amsmath}
\usepackage{graphicx}



\usepackage{babel}
\begin{document}

\title{Perturbation of FRW Spacetime in NP Formalism}

\author{S. K. Sharma and U. Khanal \\
Central Department of Physics, Tribhuvan University, Kirtipur, \textbf{Nepal}}

\date{\today}
\maketitle
\begin{abstract}
Jacobi polynomials appear to play a very important role in describing
all the spin field (s = 0, 1/2, 1, 2) perturbation of the FRW spacetime.
The formulation becomes very transparent when done in NP formalism.
All the variables are separable, and the spatial eigenfuctions turn
out to be Jacobian polynomials in different forms. In particular,
the angular ones are expressible as spin weighted spherical harmonics
which are just the spherical harmonics formed with Jacobi polynomials.
The radial eigenfuctions are also Jacobi polynomials but with unconventional
parameters. Various properties of these polynomials are used to describe
the scalar, vector and tensor modes of the perturbation. The Green's
function of the scalar perturbations and also its Lienard Wiecherte
type potentials are derived, and are shown to reduce to the familiar
ones in the limit to flat FRW case. Some of the components of the
perturbed metric tensor $h^{\mu\nu}$ have also been calculated.
\end{abstract}

\subsubsection*{PACS: 03.65.Pm, 04.20.Cv, 04.98.80.Jk}

Keywords: FRW space-time, perturbation, NP formalism, spin fields,
Jacobi polynomials.

\section{Introduction}

While doing perturbation analysis and studying structure formation
process, if we use conventional tensoral method, it is difficult to
linearize Eistein's equation. Even if it is linearized, it is not
easy to get the solutions and fully interpret them . Some of the informations
are lost in the process of linearizations. All these problems are
solved and the solutions are obtained in more natural ways if we use
Newman-Penrose (NP) formalism. In NP formalism, the unperturbed /
background quantities vanish and the equations left behind are in
terms of perturbed quantities only. So, it is more natural in analysing
the perturbation and structure formation.

The Newman-Penrose\cite{NP} formalism of projecting vectors, tensors
and spinors onto a set of null tetrad bases has proved to be an immensely
useful tool to investigate the properties of quantum fields in curved
spacetime. It has been used successfully in various black hole geometries
\cite{chandra1,chandra2,chandra3,lohia,lohia1,uk,uk-pk,uk-pk1,uk3,uk4}.
The method has also been used more\ recently\cite{mukhopadhyay}
to further study the behaviour of Dirac particles in different geometries.
As the geometry of our Universe is of the Friedmann-Robertson-Walker
type, it is important to study how electrodynamics is altered by the
real expanding Universe in contrast to that in Minkowskian spacetime.
The behaviour of the quantum fields in FRW spacetime must be indicative
of the exact nature of the geometry of our Universe, particularly
whether it is flat, closed or open. Properties of matter fields in
general, and the massive Dirac fields in particular, must have profound
consequences on the structure formation process. Propagation of electromagnetic
waves in FRW spacetime was studied by Haghighipour\cite{nader} (see
references therein), among other authors. The Dirac field was studied
in the NP formalism by Zecca\cite{zecca,zecca-montaldi} and the work
was continued further by Sharif\cite{sharif}. 

Motivated by the success of the Newman-Penrose(NP) method in investigating
the perturbation of various black hole geometries, we thought that
the application of this method to study the perturbation of the Friedmann-Robertson-Walker
spacetime may give some further insight into structure formation.
One of the authors \cite{uk1,uk2} had used this formalism to study
free Maxwell and Dirac fields in FRW space. After setting up the notations
below, we write down the perturbation equations in the next section
and solve them. Section 3 discusses the scalar perturbation, and the
Green's function as well as the Lienard-Wiechert type potential in
closed space are found. In the final section we discuss some consequences.

Writing the line element as \cite{weinberg,adler}

\begin{equation}
ds^{2}=a^{2}\left[d\eta^{2}-dr^{2}-S^{2}\left(d\theta^{2}+\sin^{2}\theta d\phi^{2}\right)\right]
\end{equation}

with $S=\frac{\sin\left(\sqrt{K}r\right)}{\sqrt{K}}$, we may devise
the null tetrad to be

$l_{\mu}=[1,-1,0,0],n_{\mu}=\frac{a^{2}}{2}[1,1,0,0],\, m_{\mu}=-\frac{aS}{\sqrt{2}}[0,0,1,i\sin\theta],$
and the complex conjugate $\bar{m}_{\mu}$ to express the directional
derivatives as

\begin{eqnarray}
D & = & l^{\mu}\partial_{\mu}=\frac{1}{a^{2}}\mathcal{D}^{-},\,\Delta=n^{\mu}\partial_{\mu}=-\frac{1}{2}\mathcal{D}^{+},\nonumber \\
\delta & = & m^{\mu}\partial_{\mu}=\frac{1}{\sqrt{2}aS}\mathcal{L}^{-},\delta^{*}=\bar{m}^{\mu}\partial_{\mu}=\frac{1}{\sqrt{2}aS}\mathcal{L}^{+}
\end{eqnarray}
 with $\mathcal{D}^{\pm}=\frac{\partial}{\partial r}\mp\frac{\partial}{\partial\eta}$
and $\mathcal{L}^{\pm}=\frac{\partial}{\partial\theta}\mp\frac{i}{\text{sin\ensuremath{\theta}}}\frac{\partial}{\partial\phi}$.
Also, the non-vanishing spin coefficients\cite{chandra} are given
by

\begin{equation}
\beta=-\alpha=\frac{\text{cot\ensuremath{\theta}}}{2\sqrt{2}S},\,\gamma=-\frac{\dot{a}}{2a},\,2\mu=\frac{\dot{a}}{a}-\frac{S^{\prime}}{S},\,-a^{2}\rho=\frac{\dot{a}}{a}+\frac{S^{\prime}}{S},
\end{equation}

where the overdot and prime denote derivatives with the conformal
time $\eta$ and $r$ respectively. 

Of the NP quantities representing the curvature of the spacetime,
all the Weyl scalers $(\Psi^{\prime}\mbox{s})$ vanish in the homogeneous
and isotropic FRW background, and the non-vanishing background Ricci
scalers are given through
\begin{eqnarray}
R & = & 6\left(\frac{\ddot{a}+Ka}{a^{6}}\right),\nonumber \\
\Phi_{00} & = & \frac{4}{a^{2}}\Phi_{11}=\frac{4}{a^{4}}\Phi_{22}=\frac{1}{a^{6}}\left(Ka^{2}+2\dot{a}^{2}-a\ddot{a}\right)=\frac{4\pi}{a^{2}}(\rho+p).
\end{eqnarray}

In the work below, we just consider the closed Universe with $K=1$,
as the other two, $K=\left(0,-1\right)$ turn out to be analytic continuation
of the same solution. 

We will also have to use the operators $\mathcal{D}_{n}^{\pm}=\mathcal{D}^{\pm}+n\frac{S^{\prime}}{S}$
and $\mathcal{L}_{n}^{\pm}=\mathcal{L}^{\pm}+n\text{cot\ensuremath{\theta}}$.

\section{Perturbation of the space-time}

Now we look at the perturbation equations. In combination with the
$\Psi$'s that vanish in the background, we should take the unperturbed
directional derivatives and spin coefficients for the first order
perturbation equations. These equations are given by

\begin{eqnarray}
\frac{1}{a^{6}S^{4}}\mathcal{D}^{-}[a^{4}S^{4}\Psi_{1}]-\frac{1}{\sqrt{2}a^{6}S^{2}}\mathcal{L}_{2}^{+}[a^{5}S\Psi_{0}] & = & -[Ricci-1],\,\,\,\,\,\,\,(a)\nonumber \\
\frac{1}{2a^{5}S}\mathcal{D}^{+}[a^{5}S\Psi_{0}]+\frac{1}{\sqrt{2}a^{5}S^{5}}\mathcal{L}_{-1}^{-}[a^{4}S^{4}\Psi_{1}] & = & -[Ricci-3]\,\,\,\,\,\,\,(e)\nonumber \\
-\frac{1}{a^{5}S^{3}}\mathcal{D}^{-}[a^{3}S^{3}\Psi_{2}]+\frac{1}{\sqrt{2}a^{5}S^{3}}\mathcal{L}_{1}^{+}[a^{4}S^{2}\Psi_{1}] & = & -[Ricci-2]\,\,\,\,\,\,\,(b)\nonumber \\
\frac{1}{2a^{4}S^{2}}\mathcal{D}^{+}[a^{4}S^{2}\Psi_{1}]+\frac{1}{\sqrt{2}a^{4}S^{4}}\mathcal{L}_{0}^{-}[a^{3}S^{3}\Psi_{2}] & = & -[Ricci-4]\,\,\,\,\,\,\,(f)\nonumber \\
\frac{1}{a^{4}S^{2}}\mathcal{D}^{-}[a^{2}S^{2}\Psi_{3}]-\frac{1}{\sqrt{2}a^{4}S^{4}}\mathcal{L}_{0}^{+}[a^{3}S^{3}\Psi_{2}] & = & -[Ricci-5]\,\,\,\,\,\,\,(c)\nonumber \\
\frac{1}{2a^{3}S^{3}}\mathcal{D}^{+}[a^{3}S^{3}\Psi_{2}]+\frac{1}{\sqrt{2}a^{3}S^{3}}\mathcal{L}_{1}^{-}[a^{2}S^{2}\Psi_{3}] & = & -[Ricci-7]\,\,\,\,\,\,\,(g)\nonumber \\
-\frac{1}{a^{3}S}\mathcal{D}^{-}[aS\Psi_{4}]+\frac{1}{\sqrt{2}a^{3}S^{5}}\mathcal{L}_{-1}^{+}[a^{2}S^{4}\Psi_{3}] & = & -[Ricci-6]\,\,\,\,\,\,\,(d)\nonumber \\
\frac{1}{2a^{2}S^{4}}\mathcal{D}^{+}[a^{2}S^{4}\Psi_{3}]+\frac{1}{\sqrt{2}a^{2}S^{2}}\mathcal{L}_{2}^{-}[aS\Psi_{4}] & = & -[Ricci-8]\,\,\,\,\,\,\,(h)\nonumber \\
\end{eqnarray}

where the explicit terms in the Ricci tensors (enclosed in square
brackets) are found on page 50 of Ref. \cite{chandra}. The letters
(a), (b), etc. correspond to those equations there.

All these can be combined in one master equation:

\begin{equation}
\sin r\,\mathcal{D}_{\mp p}^{\pm}\Phi_{p}\pm\mathcal{L}_{1\mp p}^{\mp}\Phi_{p\mp1}=U_{\mp p}^{\pm}\label{eq: master}
\end{equation}

and the spin fields are 

\begin{equation}
\Phi_{\pm2}=\begin{cases}
\frac{a^{5}}{2}S^{3}\Psi_{0},\\
2aS\Psi_{4},
\end{cases}
\end{equation}

representing the gravitational or tensorial perturbations, 

\begin{equation}
\Phi_{\pm1}=\begin{cases}
\frac{a^{4}}{\sqrt{2}}S^{3}\Psi_{1},\\
\sqrt{2}a^{2}S^{3}\Psi_{3},
\end{cases}
\end{equation}

representing the vectorial or vortical perturbations, and

\begin{equation}
\Phi_{0}=a^{3}S^{3}\Psi_{2}
\end{equation}

the scalar perturbations. The sources are given by $U^{\pm}$ which
can be identified as 

\begin{eqnarray}
U_{-2}^{+}=-a^{5}S^{4}\,\,\left[Ricci-3\right] & , & U_{-2}^{-}=-2a^{3}S^{4}\,\,\left[Ricci-6\right]\nonumber \\
U_{-1}^{+}=-\sqrt{2}a^{4}S^{4}\,\,\left[Ricci-4\right] & , & U_{-1}^{-}=-\sqrt{2}a^{4}S^{4}\,\,\left[Ricci-5\right]\nonumber \\
U_{+1}^{-}=-\frac{1}{\sqrt{2}}a^{6}S^{4}\,\,\left[Ricci-1\right] & , & U_{+1}^{+}=-2\sqrt{2}a^{2}S^{4}\,\,\left[Ricci-8\right]\nonumber \\
U_{0}^{+}=-2a^{3}S^{4}\,\,\left[Ricci-7\right] & , & U_{0}^{-}=a^{5}S^{4}\,\,\left[Ricci-2\right]\nonumber \\
\label{eq: sources}
\end{eqnarray}

Although Eq. (\ref{eq: master}) appears to be valid for massless
spin fields of all helicities, we can define the source term only
for $-2\leq p\leq2$. To decouple the spin fields in Eq. (\ref{eq: master}),
we consider the Eq. (\ref{eq: master}) with $p\rightarrow p\pm1$,
interchanging the upper and lower signs to get

\begin{equation}
S\,\mathcal{D}_{\pm(p\mp1)}^{\mp}\Phi_{p\mp1}\mp\mathcal{L}_{\pm p}^{\pm}\Phi_{p}=U_{\pm(p\mp1)}^{\mp}\label{eq: master-2}
\end{equation}

Then operating on Eq. (\ref{eq: master}) with $S\,\mathcal{D}_{\pm(p\mp1)}^{\mp}$
and Eq. (\ref{eq: master-2}) with $\mathcal{\mp L}_{\mp(p\mp1)}^{\mp}$
and adding the two, we find

\begin{eqnarray}
\left[S\,\mathcal{D}_{\mp p-1}^{\mp}\,\, S\,\mathcal{D}_{\mp p}^{\pm}+\mathcal{L}_{\mp p+1}^{\mp}\mathcal{L}_{\pm p}^{\pm}\right]\Phi_{p} & = & S\,\mathcal{D}_{\pm p-1}^{\mp}U_{\mp p}^{\pm}\nonumber \\
 &  & \mp\mathcal{L}_{\mp p+1}^{\mp}U_{\pm p-1}^{\mp}
\end{eqnarray}

The equations for the vectorial perturbations are the same as the
Maxwell equations discussed in Ref.\cite{uk1}. It is an interesting
proposition that the vortical perturbations arise from the electrodynamic
interaction between the Maxwell and Dirac fields. Simlarly, the equation
governing the scalar perturbation is the same as that of the conformally
coupled massless Klein-Gordon field. There are $10$ degrees of gauge
freedom in the problem as we have formulated, four relating to the
general co-variance of general relativity and six corresponding to
rotations of the tetrad frame as discussed in Ref. \cite{chandra}.
These conditions can best be used to simplify the source term, but
we will not be concerned with the sources in this work, and will only
discuss the eigen-modes of these perturbations.

It is clear that the angular part is separable, and we may write $\Phi_{p}=Z_{p}\,\times\, Y_{p}$.
The eigenequations for $Y_{p}$ are represented by the the spin-weighted
spherical harmonics satisfying 
\begin{eqnarray}
\mathcal{L}_{1+p}^{+}\mathcal{L}_{-p}^{-}\,_{p}Y_{l}^{m} & = & -(l-p)(l+p+1)\,_{\, p}Y_{l}^{m};\\
 & = & -L_{p}^{2}\,\,\,\,_{\, p}Y_{l}^{m}
\end{eqnarray}
the solution\cite{abram} is

\begin{equation}
_{p}Y_{l}^{m}\left(\theta,\phi\right)=Ne^{im\phi}(1-\text{cos\ensuremath{\theta}})^{\frac{m+p}{2}}(1+\text{cos\ensuremath{\theta}})^{\frac{m-p}{2}}P_{l-m}^{(m+p,m-p)},
\end{equation}

allowing us to identify the spin-weighted spherical harmonics as the
spherical harmonics formed with the Jacobi polynomial $P_{n}^{(\alpha,\beta)}$.
It can easily be established that $_{p}Y_{l}^{m}{}^{\ast}=(-1)^{m+p}{}_{-p}Y_{l}^{-m}$,
and $_{0}Y_{l}^{m}=Y_{l}^{m}$ are just the usual spherical harmonics.
These are normalized to give $\int_{p}Y_{l^{\prime}}^{m^{\prime}}{}^{\ast}{}_{p}Y_{l}^{m}\text{d\ensuremath{\Omega}}=\delta_{ll^{\prime}}\delta_{mm^{\prime}}$,
and they form a complete set in that $\sum_{l,m}\,_{p}Y_{l}^{m}(\Omega^{\prime}){}^{\ast}{}_{p}Y_{l}^{m}$($\Omega$)=$\delta$($\Omega$-$\Omega$').
Koornwinder\cite{koorn} developed an addition formula for Jacobi
polynomials, and there is a generalized addition theorem for $_{p}Y_{l}^{m}$
given in Ref.\cite{hu}, but we can devise a degenerate form for our
purpose by noting that $\delta\left(\Omega-\Omega'\right)$ can depend
only on the angle $\beta$ between the directions $\Omega\mbox{ and }\Omega'$
where 

\begin{equation}
\cos\beta=\cos\theta\cos\theta'+\sin\theta\sin\theta'\cos\left(\phi-\phi'\right).
\end{equation}

So, multiplying $\delta$($\Omega$-$\Omega$')=$\sum_{l}b_{l}\,\,_{p}Y_{l}^{-p}\left(\beta,0\right)$
on both sides by $_{p}Y_{l'}^{-p}\left(\beta,0\right)$ and integrating
over the solid angle to find $b_{l}=_{p}Y_{l}^{-p}\left(0,0\right)$,
we may write 
\begin{equation}
\sum_{m}\,_{p}Y_{l}^{m*}\left(\Omega'\right){}_{p}Y_{l}^{m}\left(\Omega\right)={}_{p}Y_{l}^{-p*}\left(0,0\right){}_{p}Y_{l}^{-p}\left(\beta,0\right).\label{2.5}
\end{equation}

For the radial eigenfunctions, we substitute $_{p}Z_{k}^{\omega}\left(\eta,r\right)=e^{-i\omega\eta}\,_{p}R_{k}^{\omega}\left(r\right)$.
With a change of variable $\cos\xi=i\cot r$, we find $S\mathcal{D}_{n}^{\pm}=i\Delta_{-n}^{\pm}=i\left(\frac{\partial}{\partial\xi}\pm\frac{\omega}{\sin\xi}+n\text{cot\ensuremath{\xi}}\right)$
which looks the same as the angular operator. Hence 

\begin{equation}
S\mathcal{D}_{\mp p-1}^{\mp}\,\, S\mathcal{D}_{\mp p}^{\pm}\,\,\,_{p}R_{k}^{\omega}=-\Delta_{\pm p-1}^{\pm}\,\,\Delta_{\pm p}^{\mp}\,\,_{p}R_{k}^{\omega}=\left(k-p\right)\left(k+p+1\right)\,_{p}R_{k}^{\omega}\label{2.6}
\end{equation}
where
\begin{eqnarray}
_{p}R_{k}^{\omega} & = & \,_{p}N_{k}^{\omega}\left(1-i\cot r\right)^{-\frac{\left(\omega+p\right)}{2}}\left(1+\cot r\right)^{-\frac{\left(\omega-p\right)}{2}}P_{\omega-k-1}^{\left(-\omega-p,-\omega+p\right)}\left(i\cot r\right)\label{2.7}
\end{eqnarray}
$N$ is the normalization constant, and the parameters of the Jacobi
polynomials are chosen to make the function regular at $r=0\mbox{ and }\pi$.
Ref. \cite{kuijla} has proved the non-Hermitian orthogonality of
Jacobi polynomials with general parameters. In our case, we find that

$N^{2}\int_{i\infty}^{-i\infty}$$\left(_{p}R_{k'}^{\omega}{}_{p}R_{k}^{\omega}\right)id\left(\cos\xi\right)=\delta_{kk'}$
gives us

\begin{equation}
N^{2}=2^{\omega-1}\left[\frac{\left(2k+1\right)\left(\omega-k-1\right)!\left(k+p\right)!\left(k-p\right)!}{\pi\left(\omega+k\right)!}\right].
\end{equation}

So $_{p}R_{k}^{\omega*}={}_{-p}R_{k}^{\omega}.$ It should be noted
that $R^{*}$ and R are not orthogonal.

Both the angular and radial functions satisfy the spin raising and
lowering operations
\begin{eqnarray}
\mathcal{L}_{\pm p}^{\pm}\,_{p}Y_{l}^{m} & = & \pm L_{\pm p}\,_{\left(p\mp1\right)}Y_{l}^{m}\mbox{ and}\nonumber \\
\Delta_{\pm p}^{\pm}\,_{p}Z_{k}^{\omega} & = & \pm K_{\pm p}\,_{\left(p\mp1\right)}Z_{k}^{\omega},\label{2.8}
\end{eqnarray}
where the eigenvalues are

$L_{p}=\sqrt{\left(l-p\right)\left(l+p+1\right)}\mbox{ }$and $K_{p}=\sqrt{\left(k-p\right)\left(k+p+1\right)}$. 

\begin{figure}
\includegraphics{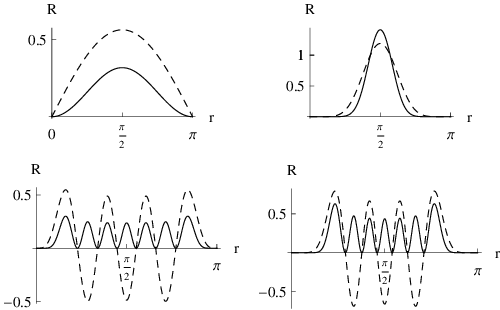} \\
\caption{The scalar radial eigen-functions. The dashed curves are the normalized
functions and the solid curves are the squares of them. The upper
two are for $n=\omega-k-1=0,$ the lowest possible state, with $k=0$
on the left, and $k=6$ on the right. The lower ones are for $n=6$,
with $k=3$ on the left and $k=8$ on the right. It is seen that the
higher $k$ states are confined towards the middle, nearer to $r=\pi/2$. }

\label{scalar} 
\end{figure}

\begin{figure}
\includegraphics[scale=1.5]{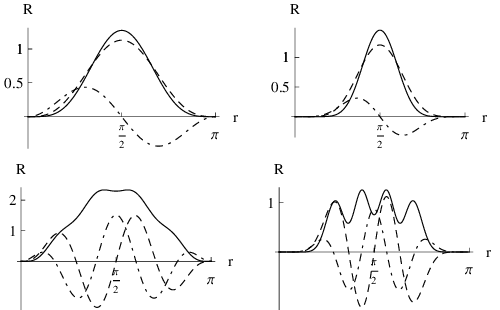} \\
\caption{The vectorial radial eigenfunctions. The dashed curves are the real
part, the dot-dashed are imaginary and the solid are the square of
the modulus. The upper two graphs are for $n=0$ for, with the lowest
state of $k=1$ on the left, and $k=4$ on the right. The lower graphs
are for $n=3$, with $k=1$ on the left and $k=6$ on the right. Again,
the higher $k\mbox{'s}$ are confined nearer to the middle. }

\label{vector} 
\end{figure}

\begin{figure}
\includegraphics{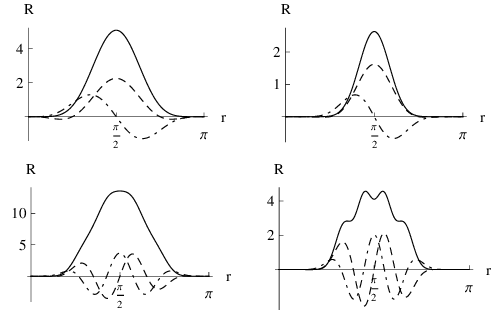} \\
\caption{Same as Fig. (\ref{vector}), with same values of $n$ for the wave
forms of the tensorial gravitational radiation. The upper left is
the lowest state with $k=2$, and the right is $k=6$. The lower ones
are $k=4\mbox{ and }10$ respectively. Higher $k$ values show similar
tendencies as the scalar and vectorial perturbations. }

\label{tensor} 
\end{figure}

In Eq. (\ref{2.8}), the upper signs lower the helicity while the
lower signs raise it. Also the eigen-values satisfy $L_{-p}\left(\mbox{or}\,\,\,\, K_{-p}\right)=L_{p-1}\left(\mbox{or}\,\,\,\, K_{p-1}\right)$.
Some of the eigen-functions representing the various modes of perturbation
are displayed in the accompanying figures. The scalar perturbations,
Fig. (\ref{scalar}), represent the density perturbations. These give
rise to the structures. Fig. (\ref{vector}) show the vectorial perturbations.
These are responsible for generating rotational or vortical effects
of the perturbations. The next mode shown in Fig. (\ref{tensor})
are the tensorial perturbation representing the gravitational radiation
content. All these modes are inter-related; e.g., any scalar density
enhancement due to gravitational collapse generates a rotational component
and induces the release of binding energy through gravitational radiation. 

In future work, it will be important to solve the equations with the
source as well. The source terms are generated by products of these
eigenfunctions. For example, the source $\Phi_{02}$ is generated
by products like $\Phi_{1}\Phi_{-1}^{*}$. 

The radial and temporal parts of the spin field wave of helicity p
are given by $_{p}Z_{k}^{\omega}={}_{p}N_{k}^{\omega}\frac{e^{-iw\eta}}{\sqrt{2\pi}}{}_{p}R_{k}^{\omega}$.
When two such waves of $p_{1}$ and $p_{2}$ are combined, the combination
must have helicity $p=p_{1}+p_{2}$; energy conservation (orthogonality
of the temporal part) requires $\omega=\omega_{1}+\omega_{2}$ and
the combined state must be linear superposition of the available momentum
states. So we can write

\begin{equation}
_{p_{1}}Z_{k_{1}}^{\omega_{1}}{}_{p_{2}}Z_{k_{2}}^{\omega_{2}}=\sum_{k=0}^{\omega_{1}+\omega_{2}-k_{1}-k_{2}-2}A_{k}\,\,\,{}_{p_{1}+p_{2}}Z_{\omega_{1}+\omega_{2}-1-k}^{\omega_{1}+\omega_{2}}
\end{equation}

The Clebsch-Gordon type coefficient $A_{k}$ can be determined to
be

\begin{equation}
A_{k}=\frac{1}{\sqrt{2\pi}}\intop_{0}^{\pi}\frac{dr}{\sin^{2}r}\,\,\,\,{}_{p_{1}+p_{2}}R_{\omega_{1}+\omega_{2}-1-k}^{\omega_{1}+\omega_{2}}\,\,\,\,{}_{p_{1}}R_{k_{1}\,\,\,}^{\omega_{1}}{}_{p_{2}}R_{k_{2}}^{\omega_{2}}\label{eq: ak}
\end{equation}

Triple integrals for conventional Jacobi polynomials have been considered
in the literatures \cite{alisauskaus} by using group theoritical
method. Here, we derive the Clebsch-Gordan type coefficients by direct
integration. So it is worthwhile to consider the product $\Phi_{p_{1}}\Phi_{p_{2}}=Z_{p_{1}}Z_{p_{2}}Y_{p_{1}}Y_{p_{2}}$.
Clebsch-Gordan expansion of the product of the angular function is
well known: 

\begin{eqnarray}
 & _{p_{1}}Y_{l_{1}}^{m_{1}}\,\,\,{}_{p_{2}}Y_{l_{2}}^{m_{2}}\nonumber \\
= & \sum_{l}\left[\frac{\left(2l_{1}+1\right)\left(2l_{2}+1\right)}{4\pi\left(2l+1\right)}\right]^{1/2}{}_{p}Y_{l}^{m}\left\langle l_{1},\,\, l_{2};\,\, m_{1},\,\, m_{2}|l,\, m\right\rangle \left\langle l_{1},\,\, l_{2};\,\,-p_{1},\,\,-p_{2}|l,\,-p\right\rangle ,
\end{eqnarray}

where $\left\langle l_{1},\,\, l_{2};\,\, m_{1},\,\, m_{2}|l,\, m\right\rangle $~~
is the Clebsch-Gordan coefficient, $m=m_{1}+m_{2}$~~, $p=p_{1}+p_{2}$,
$\left|l_{1}-l_{2}\right|\leq l\leq l_{1}+l_{2}$. Here we will develop
a similar expansion for z.

With $z=i\cot r$, and $_{p}R_{k}^{\omega}={}_{p}N_{k}^{\omega}\left(1-z\right)^{-\left(\omega+p\right)/2}\left(1+z\right)^{-\left(\omega-p\right)/2}P_{\omega-k-1}^{\left(-\omega-p,-\omega+p\right)}\left(z\right)$,
substituting the Rodrigue's formula for the Jacobi polynomial in the
first part of the integrand of Eq. (\ref{eq: ak}) and integrating
by parts k times (with the integrated part vanishing at the boundary),
we find

\begin{align}
A_{k} & =\frac{_{p}N_{k}^{\omega}.{}_{p_{1}}N_{k_{1}}^{\omega_{1}}.{}_{p_{2}}N_{k_{2}}^{\omega_{2}}}{\sqrt{2\pi}}\frac{1}{2^{k}k!}\int_{i\infty}^{-i\infty}d\left(iz\right)\left(1-z\right)^{-\left(\omega_{1}+\omega_{2}-k+p_{1}+p_{2}\right)}\nonumber \\
 & \left(1+z\right)^{-\left(\omega_{1}+\omega_{2}-k-p_{1}-p_{2}\right)}\frac{d^{k}}{dz^{k}}\left[P_{\omega_{1}-k_{1}-1}^{\left(-\omega_{1}-p_{1},-\omega_{1}+p_{1}\right)}\left(z\right)\,\,\, P_{\omega_{2}-k_{2}-1}^{\left(-\omega_{2}-p_{2},-\omega_{2}+p_{2}\right)}\left(z\right)\right]
\end{align}

Using $\frac{d^{m}}{dz^{m}}P_{n}^{\left(-\alpha,\,-\beta\right)}\left(z\right)=\left(-1\right)^{m}\frac{m!}{2^{m}}\left(\begin{array}{c}
\alpha+\beta-n-1\\
m
\end{array}\right)P_{n-m}^{\left(-\alpha+m,\,-\beta+m\right)}\left(z\right)$ and the Leibniz rule, we have

\begin{align}
 & \frac{1}{k!}\frac{d^{k}}{dz^{k}}\left[P_{\omega_{1}-k_{1}-1}^{\left(-\omega_{1}-p_{1},-\omega_{1}+p_{1}\right)}\left(z\right)\,\,\, P_{\omega_{2}-k_{2}-1}^{\left(-\omega_{2}-p_{2},-\omega_{2}+p_{2}\right)}\left(z\right)\right]\nonumber \\
 & =\frac{\left(-1\right)^{k}}{2^{k}}\sum_{m=0}^{k}\left(\begin{array}{c}
\omega_{1}+k_{1}\\
k-m
\end{array}\right)\left(\begin{array}{c}
\omega_{2}+k_{2}\\
m
\end{array}\right)P_{\omega_{1}-k_{1}-1-k+m}^{\left(-\omega_{1}+k-m+p_{1},-\omega_{1}+k-m-p_{1}\right)}\left(z\right)\,\,\, P_{\omega_{2}-k_{2}-1-m}^{\left(-\omega_{2}+m-p_{2},-\omega_{2}+m+p_{2}\right)}\left(z\right)\nonumber \\
 & =\left(-\frac{1}{2}\right)^{\omega_{1}+\omega_{2}-k+p_{1}+p_{2}}\frac{\left(1-z\right)^{-\left(\omega_{1}+\omega_{2}-k+p_{1}+p_{2}\right)}\left(1+z\right)^{-\left(\omega_{1}+\omega_{2}-k-p_{1}-p_{2}\right)}}{\left(\omega_{1}+\omega_{2}-k_{1}-k_{2}-k-z\right)!}\nonumber \\
 & \sum_{m=0}^{k}\left(\begin{array}{c}
\omega_{1}+k_{1}\\
k-m
\end{array}\right)\left(\begin{array}{c}
\omega_{2}+k_{2}\\
m
\end{array}\right)\left[\left(\frac{d}{dz}\right)^{\omega_{1}-k_{1}-1-k-m}\left(1-z\right)^{-\left(k_{1}+1+p_{1}\right)}\left(1+z\right)^{-\left(\omega_{1}+1-p_{1}\right)}\right]\nonumber \\
 & \left(\begin{array}{c}
\omega_{1}+\omega_{2}-k_{1}-k_{2}-k-2\\
\omega_{2}-k_{2}-1-m
\end{array}\right)\left[\left(\frac{d}{dz}\right)^{\omega_{2}-k_{2}-1-m}\left(1-z\right)^{-\left(k_{2}+1-p_{2}\right)}\left(1+z\right)^{-\left(k_{2}+1+p_{2}\right)}\right]
\end{align}

Substituting these expressions and further integration by parts gives

\begin{eqnarray}
A_{k} & = & \frac{_{p}N_{k}^{\omega}.{}_{p_{1}}N_{k_{1}}^{\omega_{1}}.{}_{p_{2}}N_{k_{2}}^{\omega_{2}}}{\sqrt{2\pi}}\frac{\left(-1\right)^{\omega_{2}-k_{2}-k-1}}{2^{\omega_{1}+\omega_{2}-k_{1}-k_{2}+k+2}\left(\omega_{1}+\omega_{2}-k_{1}-k_{2}-k-2\right)!}\nonumber \\
 &  & \sum_{m=0}^{k}\left(-1\right)^{m}\left(\begin{array}{c}
\omega_{1}+k_{1}\\
k-m
\end{array}\right)\left(\begin{array}{c}
\omega_{2}+k_{2}\\
m
\end{array}\right)\left(\begin{array}{c}
\omega_{1}+\omega_{2}-k_{1}-k_{2}-k-2\\
\omega_{2}-k_{2}-1-m
\end{array}\right)\nonumber \\
 &  & \int_{i\infty}^{-i\infty}d\left(iz\right)\left(1-z\right)^{-\left(k_{1}+1+p_{1}\right)}\left(1+z\right)^{-\left(\omega_{1}+1-p_{1}\right)}\nonumber \\
 &  & \left(\frac{d}{dz}\right)^{\omega_{1}+\omega_{2}-k_{1}-k_{2}-k-2}\left(1-z\right)^{-\left(k_{2}+1+p_{2}\right)}\left(1+z\right)^{-\left(k_{2}+1-p_{2}\right)}\nonumber \\
 & = & \sum_{q=0}^{\omega_{1}+\omega_{2}-k_{1}-k_{2}-k-2}\left(-1\right)^{q}\left(\begin{array}{c}
\omega_{1}+\omega_{2}-k_{1}-k-2+p_{2}-q\\
k_{2}+p_{2}
\end{array}\right)\left(\begin{array}{c}
k_{2}-p_{2}+q\\
k_{2}-p_{2}
\end{array}\right)\nonumber \\
 &  & \int_{i\infty}^{-i\infty}d\left(iz\right)\left(1-z\right)^{-\left(\omega_{1}+\omega_{2}-k-q+p_{1}+p_{2}\right)}\left(1+z\right)^{-\left(k_{1}+k_{2}+2+q-p_{1}-p_{2}\right)}\label{eq:2.9}
\end{eqnarray}

Since $\left(1\pm z\right)=\left(1\pm i\,\cot r\right)=\pm i\frac{e^{\mp ir}}{\sin r}$,
we have

\begin{eqnarray}
 & \int_{i\infty}^{-i\infty}d\left(iz\right)\left(1-z\right)^{-\left(\omega_{1}+\omega_{2}-k-q+p_{1}+p_{2}\right)}\left(1+z\right)^{-\left(\omega_{1}+k_{2}+2+q-p_{1}-p_{2}\right)}\nonumber \\
= & i^{\left(\omega_{1}+\omega_{2}-k_{1}-k_{2}-2-k-2q+2p_{1}+2p_{2}\right)}\int_{0}^{\pi}dr\,\, e^{-ir\left(\omega_{1}+\omega_{2}-k_{1}-k_{2}-2-k-2q+2p_{1}+2p_{2}\right)}\left(\sin r\right)^{\omega_{1}+\omega_{2}-k_{1}-k_{2}-k}\nonumber \\
= & \frac{\pi}{2^{\omega_{1}+\omega_{2}-k_{1}-k_{2}-k}}\,\,\frac{1}{\left(\omega_{1}+\omega_{2}+k_{1}+k_{2}-1-k\right)}\,\,\frac{1}{B\left(\omega_{1}+\omega_{2}-q+p_{1}+p_{2},\,\, k_{1}+k_{2}+k+q-p_{1}-p_{2}\right)}\nonumber \\
= & \frac{\pi}{2^{\omega_{1}+\omega_{2}-k_{1}-k_{2}-k}}\left(\begin{array}{c}
\omega_{1}+\omega_{2}+k_{1}+k_{2}+k-2\\
\omega_{1}+\omega_{2}+p_{1}+p_{2}-q
\end{array}\right)\label{eq: 2.10}
\end{eqnarray}

So, combining Equations (\ref{eq:2.9}) and (\ref{eq: 2.10}), we
get 

\begin{eqnarray}
A_{k} & = & \frac{_{p}N_{k}^{\omega}.{}_{p_{1}}N_{k_{1}}^{\omega_{1}}.{}_{p_{2}}N_{k_{2}}^{\omega_{2}}}{\sqrt{2\pi}}\frac{\left(-1\right)^{\omega_{2}-k_{2}-1}}{2^{2\left(\omega_{1}+\omega_{2}+1\right)}}\nonumber \\
 &  & \sum_{m=0}^{k}\left(-1\right)^{m}\left(\begin{array}{c}
\omega_{1}+k_{1}\\
k-m
\end{array}\right)\left(\begin{array}{c}
\omega_{2}+k_{2}\\
m
\end{array}\right)\left(\begin{array}{c}
\omega_{1}+\omega_{2}-k_{1}-k_{2}-k-2\\
\omega_{2}-k_{2}-1-m
\end{array}\right)\nonumber \\
 &  & \sum_{q=0}^{\omega_{1}+\omega_{2}-k_{1}-k_{2}-k-2}\left(-1\right)^{q}\left(\begin{array}{c}
\omega_{1}+\omega_{2}-k_{1}-k-2+p_{2}-q\\
k_{2}+p_{2}
\end{array}\right)\left(\begin{array}{c}
k_{2}-p_{2}+q\\
k_{2}-p_{2}
\end{array}\right)\nonumber \\
 &  & \frac{\pi}{2^{\left(\omega_{1}+\omega_{2}\right)}}\,\,\left(\begin{array}{c}
\omega_{1}+\omega_{2}+k_{1}+k_{2}+k-2\\
\omega_{1}+\omega_{2}+p_{1}+p_{2}-q
\end{array}\right)
\end{eqnarray}

\section{Scalar Perturbation}

Let us look at the scalar perturbation in some greater detail. To
solve the inhomogeneous equation with source, we can work out the
Green's function by writing Eq. (\ref{eq: master-2}) for $p=0$ with
the delta function source\cite{Sharma}: 
\begin{equation}
\left[\frac{\partial}{\partial r^{2}}-\frac{\partial}{\partial\eta^{2}}+\frac{1}{\sin^{2}r}\mathcal{L}_{1}^{+}\mathcal{L}_{0}^{-}\right]G\left(\eta,\,\, r,\,\,\,\theta,\,\,\phi\right)=\frac{1}{\sin^{2}r}\delta\left(\eta-\eta'\right)\delta\left(r-r'\right)\delta\left(\theta-\theta'\right)\delta\left(\phi-\phi'\right).\label{3.1}
\end{equation}
The temporal eigen-functions are $e^{-i\omega\eta}$, the angular
are the spherical harmonics $_{0}Y_{l}^{m}=Y_{l}^{m}$ and the radial
ones are the normalized and appropriately weighted Gegenbauer polynomial
$R_{k}=\sqrt{N}\sin^{l+1}r\, C_{n}^{l+1}$. Now we can immediately
write down the eigen-function expansion of the Green's function \cite{mathew}
as 
\begin{eqnarray}
G & = & \frac{1}{2\pi}\sum_{k,l,m}\int_{-\infty}^{\infty}\frac{d\omega}{\omega^{2}-\left(k+l+1\right)^{2}}e^{-i\omega\left(\eta-\eta'\right)}Y_{l}^{m*}\left(\Omega'\right)Y_{l}^{m}\left(\Omega\right)\nonumber \\
 &  & \times N\left(\sin r\,\sin r'\right){}^{l+1}C_{k}^{l+1}\left(\cos r\right)C_{k}^{l+1}\left(\cos r'\right)\label{3.2}
\end{eqnarray}
Firstly, the integral over $\omega$ can be done by the method of
residues to give

\begin{equation}
\int_{-\infty}^{\infty}\frac{d\omega}{\omega^{2}-\left(k+l+1\right)^{2}}e^{-i\omega\left(\eta-\eta'\right)}=-2\pi\sin\left[\omega\left(k+l+1\right)\left(\eta-\eta'\right)\right].
\end{equation}

Next, the addition theorem for spherical harmonics can be used to
write

\begin{equation}
\sum_{m}Y_{l}^{m*}\left(\Omega'\right)Y_{l}^{m}\left(\Omega\right)=\frac{2l+1}{4\pi}C_{l}^{\left(1/2\right)}\left(\cos\beta\right)
\end{equation}

where$\beta$ is the angle between the directions $\Omega\mbox{ and }\Omega'$.
Then we are left with 
\begin{eqnarray}
G & = & -\frac{1}{2\pi^{2}}\sum_{k=0}^{\infty}\sin\left[\left(k+1\right)\left(\eta-\eta'\right)\right]\sum_{l=0}^{k}\frac{2^{2l}\left(2l+1\right)\, l!^{2}\,\left(k-l\right)!}{\left(k+l+1\right)!}\nonumber \\
 &  & \times\left(\sin r\,\sin r'\right)^{l}C_{k-l}^{\left(l+1\right)}\left(\cos r\right)C_{k-l}^{\left(l+1\right)}\left(\cos r'\right)
\end{eqnarray}
Here, we may use the addition theorem for Gegenbauer polynomials\cite{grads}
to replace the second summation with 

\begin{equation}
C_{n}^{\left(1\right)}\left(\cos\rho\right)=\frac{\sin\left[\left(n+1\right)\rho\right]}{\sin\rho},
\end{equation}

where $\cos\rho=\cos r\cos r'+\sin r\sin r'\cos\beta$ . Thus, 
\begin{eqnarray}
G & = & -\frac{1}{4\pi^{2}\sin\rho}\sum_{k=1}^{\infty}\left\{ \cos k\left[\rho-\left(\eta-\eta'\right)\right]-\cos k\left[\rho+\left(\eta-\eta'\right)\right]\right\} \nonumber \\
 & = & -\frac{1}{4\pi^{2}\sin\rho}\left\{ \delta\left[\rho-\left(\eta-\eta'\right)\right]-\delta\left[\rho+\left(\eta-\eta'\right)\right]\right\} .\label{3.3}
\end{eqnarray}
The first term within the braces is the retarded, and the second is
advanced, Green's function. In the limit to the flat case $\left(r,r'\right)\rightarrow0$,
we find $\sin\rho\rightarrow\rho$, and $\cos\rho\rightarrow1-\frac{\rho^{2}}{2}=\left(1-\frac{r^{2}}{2}\right)\left(1-\frac{r'^{2}}{2}\right)+rr'\cos\beta\Rightarrow\rho^{2}\rightarrow\left|\vec{r}-\vec{r'}\right|^{2}$;
hence we recover the familiar Green's function from classical electrodynamics.
With Eq. (\ref{3.3}), we can consider the potential $s\left(\eta',r',\theta',\phi'\right)=\delta\left[r'-\zeta\left(\eta'\right)\right]\delta\left[\theta'-\theta_{\zeta}\left(\eta'\right)\right]\delta\left[\phi'-\phi_{\zeta}\left(\eta'\right)\right]$
of a point scalar perturbation moving along the trajectory $\left(\zeta\left(\eta\right),\theta_{\zeta}\left(\eta\right),\phi_{\zeta}\left(\eta\right)\right)$
to find a retarded Lienard-Wiechert type potential. Thus,

\begin{eqnarray}
\Phi_{0}\left(\rho,\eta\right) & = & \int G\, s\, dr'd\theta'd\phi'd\eta'\nonumber \\
 & = & -\frac{1}{4\pi}\int d\eta'\frac{1}{\sin\left(\rho\left(\eta'\right)\right)}\delta\left[\rho\left(\eta'\right)-\left(\eta-\eta'\right)\right].
\end{eqnarray}

The integral over $\eta'$can now be easily done to give 
\begin{equation}
\Phi_{0}\left(\rho,\eta\right)=-\left.\frac{1}{4\pi\sin\left[\rho\left(\eta'\right)\right]\left[1+\frac{d}{d\eta'}\rho\left(\eta'\right)\right]}\right|_{\eta'=\eta-\rho\left(\eta'\right)},\label{3.4}
\end{equation}
where the variables are to be evaluated at the retarded time $\eta'=\eta-\rho\left(\eta'\right)$;
here,

\begin{equation}
\cos\left[\rho\left(\eta'\right)\right]=\cos r\,\cos\left[r'\left(\eta'\right)\right]+\sin r\,\sin\left[r'\left(\eta'\right)\right]\cos\left[\beta\left(\eta'\right)\right]
\end{equation}

and

\begin{equation}
\cos\left[\beta\left(\eta'\right)\right]=\cos\theta\,\cos\left[\theta'\left(\eta'\right)\right]+\sin\theta\,\sin\left[\theta'\left(\eta'\right)\right]\cos\left[\phi-\phi\left(\eta'\right)\right],
\end{equation}

with the primed co-ordinates being the location of the source that
produce the potential at the un-primed location of the observer; also
the right hand side is to be evaluated at the retarded time. Without
loss of generality, we may simplify the notation by locating the observer
at the origin, in which case $\cos\left[\rho\left(\eta'\right)\right]=\cos\left[r'\left(\eta'\right)\right]$.
Again it can be easily checked that Eq. (\ref{3.4}) reduces to the
familiar form in the limit $r\rightarrow\mbox{small}$.

\section{Relation to metric perturbation: application to shearing}

The perturbed tetrad is written as a superposition of the unperturbed
ones \textit{viz}., $e_{a}^{(1)}=A_{a}^{k}\,\, e_{k}$ where $A_{a}^{k}$
are first order in perturbation and $e_{k}=\left[e_{1},\,\, e_{2},\,\, e_{3},\,\, e_{4}\right]=\left(l,\,\, n,\,\, m,\,\,\bar{m}\right)$
and the superscript $(1)$ indicates first order in perturbation.
$A_{1}^{1},\,\, A_{2}^{1},\,\, A_{1}^{2},\,\, A_{2}^{2}$ are real
and others complex with the interchange of indices $3\leftrightarrow4$
given the complex cojugate. Hence 16 real functions are required to
specify all $A_{a}^{k}$. These are subject to 10 gauge freedoms,
4 of general covariance and 6 of tetrad rotations.

The Ricci identities provide the equations satisfied by the spin coefficients.
In particular, the one that relates the shear $\sigma$ to the gravitational
radiation is (\cite{chandra}, Eq. 8.310 b)

\begin{equation}
aS\,\, S\mathcal{D}_{1}^{-}a^{2}S\sigma-\mathcal{L}_{-1}^{-}\,\, a^{4}S^{2}k/\sqrt{2}=2\,\,\Phi_{2}
\end{equation}

Using the boost and spin weight raising and lowering operators ($\mathcal{L}$~~and~~$S\mathcal{D}$),
we are just able to read off the solutions

\begin{equation}
a^{3}S^{2}\sigma=S\mathcal{D}_{-2}^{+}\frac{2\Phi_{2}}{K_{1}^{2}}\label{eq: sol-1}
\end{equation}

and

\begin{equation}
a^{4}S\frac{\kappa}{\sqrt{2}}=\left[\mathcal{D}_{0}^{-}\ln aS\right]S\mathcal{D}_{-2}^{+}\,\,\mathcal{L}_{2}^{+}\frac{2\Phi_{2}}{K_{1}^{2}\,\, L_{1}^{2}}\label{eq: sol-2}
\end{equation}

To relate the spin coefficients to $A_{a}^{k}$ we need to linearize
the commutation relation $c_{ab}^{k}e_{k}=\left[e_{a},\,\, e_{b}\right]$
to first order in perturbation by writing $c_{ab}^{k}=c_{ab}^{(0)\, k}+c_{ab}^{(1)\,\, k}$
to get 

\begin{equation}
c_{ab}^{(1)\,\, k}=\hat{e}_{a\,\,}A_{b}^{k}-\hat{e}_{b\,\,}A_{a}^{k}+A_{a}^{j}\,\, c_{jb}^{(0)\, k}+A_{b}^{\, j}\,\, c_{aj}^{(0)\, k}-A_{j}^{\, k}\,\, c_{ab}^{(0)\, j}
\end{equation}

where $\hat{e}_{a}$ are the directional derivatives. For our purpose
the two important ones are 

\begin{equation}
c_{13}^{(1)\,\,4}=\sigma=\frac{1}{a^{2}}\mathcal{D}_{0}^{-}A_{3}^{4}-\frac{1}{\sqrt{2}aS}\mathcal{L}_{-1\,\,}^{-}A_{1}^{4}\label{eq: sol-3}
\end{equation}

and

\begin{equation}
c_{31}^{(1)\,\,2}=\kappa=\frac{1}{\sqrt{2}aS}\mathcal{\, L}_{0\,\,}^{-}A_{1}^{2}-\frac{1}{a^{3}}\mathcal{D}_{0}^{-}\, aA_{3}^{2}\label{eq: sol-4}
\end{equation}

Now Eqs. (\ref{eq: sol-1}) and (\ref{eq: sol-3}) can be used together
to get 

\begin{equation}
aSA_{3}^{4}=S\mathcal{D}_{-1}^{+}\,\, S\mathcal{D}_{-2}^{+}\,\,\frac{2\Phi_{2}}{K_{0}^{2}K_{1}^{2}}
\end{equation}

and

\begin{equation}
a^{2}\frac{A_{1}^{4}}{\sqrt{2}}=\left[\mathcal{D}_{0}^{-}\ln aS\right]\,\, S\mathcal{D}_{-1}^{+}\,\, S\mathcal{D}_{-2}^{+}\,\,\mathcal{L}_{2\,\,}^{+}\frac{2\Phi_{2}}{K_{0}^{2}\, K_{1}^{2}\, L_{1}^{2}}
\end{equation}

With Eqs. (\ref{eq: sol-2}) and (\ref{eq: sol-4}), we find $A_{3}^{2}=-A_{1}^{4}$
and 

\begin{equation}
\frac{a^{2}A_{1}^{2}}{2}=Re\left[\mathcal{D}_{0}^{-}\left(\frac{S}{a}\mathcal{D}_{0}^{-}\ln aS\right)\right]S\mathcal{D}_{-1}^{+}\,\, S\mathcal{D}_{-2}^{+}\,\,\mathcal{L}_{1\,\,}^{+}\mathcal{L}_{2\,\,}^{+}\frac{2\Phi_{2}}{K_{0}^{2}\, K_{1}^{2}\, L_{0}^{2}\, L_{1}^{2}}
\end{equation}

Next, using similar Ricci identities for $\lambda$and $\nu$, and
the related structure constants, we find 

\begin{equation}
aSA_{4}^{3}=S\mathcal{D}_{-1}^{-}\,\, S\mathcal{D}_{-2}^{-}\,\,\frac{2\Phi_{-2}}{K_{0}^{2}K_{1}^{2}}
\end{equation}

\begin{equation}
-\sqrt{2}\, A_{2}^{3}=\sqrt{2}A_{4}^{1}=\left(\mathcal{D}_{0}^{+}\ln aS\right)\,\, S\mathcal{D}_{-1}^{-}\,\, S\mathcal{D}_{-2}^{-}\,\,\mathcal{L}_{2\,\,}^{-}\frac{2\Phi_{-2}}{K_{0}^{2}\, K_{1}^{2}\, L_{1}^{2}}
\end{equation}

and

\begin{equation}
\frac{2\, A_{2}^{1}}{a^{2}}=Re\left[\mathcal{D}_{0}^{+}\left(\frac{S}{a}\mathcal{D}_{0}^{+}\ln aS\right)\right]S\mathcal{D}_{-1}^{-}\,\, S\mathcal{D}_{-2}^{-}\,\,\mathcal{L}_{1\,\,}^{-}\mathcal{L}_{2\,\,}^{-}\frac{2\Phi_{-2}}{K_{0}^{2}\, K_{1}^{2}\, L_{0}^{2}\, L_{1}^{2}}
\end{equation}

With these solutions, we can now determine some of the components
of the perturbed metric tensor given by

\begin{equation}
h^{\mu\nu}=g^{\mu\nu}-g^{(0)\mu\nu}=\left(A_{a}^{k}\,\, e_{k}^{\mu}\,\, e_{b}^{\nu}+A_{b}^{k}\,\, e_{a}^{\mu}\,\, e_{k}^{\nu}\right)\eta^{ab}
\end{equation}

For example, 

\begin{equation}
h^{\theta\varphi}=\frac{i}{a^{2}S^{2}\sin\theta}\left(A_{3}^{4}-A_{4}^{3}\right)=\frac{i}{\left(aS\right)^{3}\sin\theta}\, S\mathcal{D}_{-1}^{+}\,\, S\mathcal{D}_{-2}^{+}\frac{2}{K_{0}^{2}\, K_{1}^{2}}\left(\Phi_{2}-\Phi_{2}^{*}\right)
\end{equation}

relates the gravitational radiation through shearing to the ($\theta\varphi$)
component of the metric tensor. It is well known that the universe
contains a background gravitational radiation at a temperatureod \textasciitilde{}
0.91K \cite{Kolb-Turner}. It is clear that this produces a shearing
which will distort the geometry of space time. We further find that
$h^{\eta\varphi}=0$. Similarly, 

\begin{eqnarray}
h^{\eta r} & = & \frac{2}{a^{2}K_{0}^{2}\, K_{1}^{2}\, L_{0}^{2}\, L_{1}^{2}}\,[\mathcal{D}_{0}^{+}\left(\frac{S}{a}\mathcal{D}_{0}^{+}\ln aS\right)\, S\mathcal{D}_{-1}^{-}\,\, S\mathcal{D}_{-2}^{-}\,\,\mathcal{L}_{1\,\,}^{-}\mathcal{L}_{2\,\,}^{-}\Phi_{-2}\nonumber \\
 & - & \mathcal{D}_{0}^{-}\left(\frac{S}{a}\mathcal{D}_{0}^{-}\ln aS\right)\, S\mathcal{D}_{-1}^{+}\,\, S\mathcal{D}_{-2}^{+}\,\,\mathcal{L}_{1\,\,}^{+}\mathcal{L}_{2\,\,}^{+}\Phi_{2}]
\end{eqnarray}

which will generate a perturbation in the radial velocity of matter.
There also appears a peculiar transverse motion generated by 

\begin{equation}
h^{\eta\theta}=\frac{1}{\sqrt{2}aS}\left[\frac{2}{a^{2}}\left(A_{2}^{3}+A_{2}^{4}\right)+\frac{1}{2}\left(A_{1}^{3}+A_{1}^{4}\right)\right]
\end{equation}

We are able to solve for 12 out of 16 real functions required to describe
the perturbed metric without using gauge conditions, leaving the four
real functions $A_{1}^{1},\,\,\,\, A_{2}^{2}$ and $A_{3}^{3}$ to
be determined. For these, we will have to include the matter producing
the perturbation. In other works, we have solved the Dirac\cite{dhungel-khanal}
as well as Maxwell equations. The matter sources are related to the
spin coefficients by other Ricci identities like

\begin{eqnarray}
\mathcal{D}_{0}^{-}a^{2}S^{2}\rho^{(1)} & - & \frac{1}{a}\left(\mathcal{D}_{0}^{-}\ln aS\right)\,\, S\mathcal{D}_{-2}^{+}\,\,\mathcal{L}_{1\,\,}^{+}\mathcal{L}_{2\,\,}^{+}\frac{2\Phi_{2}}{K_{1}^{2}\, L_{1}^{2}}\nonumber \\
 & = & a^{4}S^{2}\left[\left(\rho^{(1)}\right)^{2}+\left|\sigma\right|^{2}+\Phi_{00}^{(1)}-l^{(1)\mu}\partial_{\mu}\rho^{(0)}\right]\nonumber \\
 & + & S^{2}\left(\mathcal{D}_{0}^{-}\ln aS\right)\left[\mathcal{D}_{0}^{-}\,\,\frac{a^{2}A_{1}^{2}}{2}+\mathcal{D}_{0}^{-}\, A_{2}^{2}\right]
\end{eqnarray}

which is used to derive the optical theorem. In the perturbing source
$\Phi_{00}^{(1)}$, we have to include the energy-momentum of the
Maxwell, Dirac, their interactions, etc. It should be noted that the
RHS of last equation mostly represent second order quantities in the
perturbations. The real part of $\rho^{(1)}$ is the compression while
its imaginary part is the rotation.

\section{Conclusions}

In this work we have written the equations of perturbation of the
FRW space-time in Newman-Penrose formalism. We find that the tensorial,
vectorial and scalar perturbations decompose into forms that reveal
the spin content very transparently. The eigenfunctions of all these
modes of perturbation can be solved in terms of appropriate spin-weighted
spherical harmonics and radial functions of similar spin content. 

The angular solutions are the well known spin weighted spherical harmonics
which are really the spherical haromonics formed with Jacobi polynomials.
The radial eigenfunctins also turn out to be Jacobi polynomials but
with unconventional parameters. We have proved the non-hermitian orthogonality
of these radial functions. We have also developed a Clebsch-Gordan
type of expansion for the product of two radial functions. These will
be used in work under preparation to describe the sources.

The scalar mode gives rise to the density perturbations that are responsible
for the large scale structures. We are able to solve for the Green's
function of this mode, and the result reduces to the familiar form
from classical electrodynamics in the limit to flat space. With the
Green's function, we should be able to solve for the potential driving
the scalar perturbations. Also, the scalar perturbation $\Psi_{2}=\frac{\Phi_{0}}{a^{3}\sin^{3}r}$
may be interpreted to represent the Newtonian potential, as in the
Schwarzscild spacetime $\Psi_{2}=-\frac{M}{r^{3}}$. 

The vectorial perturbation equations are the same as Maxwell equations.
So it is worthwhile to investigate the possibility that the vortical
modes and the rotational effects on the large, and even small, scale
structures, which could have their origin in electrodynamic interactions
in very early Universe. Although the free Dirac and Maxwell equations
have been solved in another work \cite{Sharma}, it is also necessary
to incorporate the quantum corrections into the model. 

In this initial formulation of the perturbations in FRW, we have been
able to determine the shearing produced by the tensorial perturbation
representing the gravitational radiation content of the Universe.
It is enticing to interpret the Planck anomally as due to such shearing.
We have shown that all known spin field perturbations can be represented
by the spin and boost weight functions, which are just the analytically
continued Jacobi polynomials. All the known massive particles are
fermionic, so study of the Dirac current and energy-momentum as well
as their interaction with the Maxwell field will reveal many features
of structures. These and similar problems will be taken up in future
work. 

\newpage{}



\begin{thebibliography}{References}
\bibitem{NP}E. T. Newman and R. Penrose, \textit{J. Math. Phys.},
\textbf{3}, 566 (1962).

\bibitem{chandra1}S. Chandrasekhar, \textit{Proc. Royal Soc. of London
A: } \textbf{343, }289 (1975).

\bibitem{chandra2}S. Chandrasekhar, \textit{Proc. Royal Soc. of London
A: } \textbf{345}, 185 (1975).

\bibitem{chandra3}S. Chandrasekhar, \textit{Proc. Royal Soc. of London
A: } \textbf{348}, 39 (1976).

\bibitem{lohia}D. Lohia and N. Panchapakesan, \textit{J. Phys. A}:
\textbf{11},1963 (1978).

\bibitem{lohia1}D. Lohia and N. Panchapakesan, \textit{J. Phys. A}:
\textbf{12}, 533 (1979).

\bibitem{uk}U. Khanal and N. Panchapakesan, \textit{Phys. Rev. D}:
\textbf{24}, 829 (1981).

\bibitem{uk-pk}U. Khanal and N. Panchapakesan, \textit{Phys. Rev.
D}:\textit{ }\textbf{24}, 835, (1981).

\bibitem{uk-pk1}U. Khanal and N. Panchapakesan, \textit{\ Annals
of Phys., }\textbf{138}, 260 (1982).

\bibitem{uk3} U. Khanal, \textit{Phys. Rev. D}: \textbf{28}, 1291,
(1983).

\bibitem{uk4} U. Khanal, \textit{Phys. Rev. D}: \textbf{32}, 879
(1985).

\bibitem{mukhopadhyay}Banibrata Mukhopadhyay and Naresh Dadhich,
\textit{Class. Quantum Grav.} \textbf{21}, 3621 (2004), and references
therein.

\bibitem{nader}Nader Haghighipour, arxiv: gr-qc/0405140 (2004); \textit{Gen.
Relativ. Gravit.} \textbf{37}, 327 (2005).

\bibitem{zecca}A. Zecca, \textit{J. Math. Phys.} \textbf{37}, 874
(1996).

\bibitem{zecca-montaldi} E. Montaldi and A. Zecca, \textit{Int. J.
Theo. Phys.}, \textbf{37}, 995, (1998).

\bibitem{sharif}M. Sharif, \textit{Chin. J. Phys. }\textbf{40}, 526,
(2002); arxiv: gr-qc/0401065 (2004).

\bibitem{uk1}U. Khanal, \textit{Class. Quantum. Grav.} \textbf{23}
4353 (2006).

\bibitem{uk2}U. Khanal, ICTP preprint IC/2006/136P.

\bibitem{weinberg}S. Weinberg, \textit{Gravitation and Cosmology:
Principles and Applications of the General Theory of Relativity} (Wiley,
New York, 1972) p. 412

\bibitem{adler}R. Adler, M. Bazin and M. Schiffer, \textit{Introduction
to General Relativity} (McGraw-Hill, New York, 1975) p. 409

\bibitem{chandra}S. Chandrasekhar, \textit{The Mathematical Theory
of Black Holes,} Oxford University Press, New York (1983).

\bibitem{abram}M. Abramowitz and I. A. Stegun, \textit{Handbook of
Mathematical Functions}, Dover New York (1976).

\bibitem{koorn}T. Koornwinder, \textit{SIAM J. Appl. Math.} \textbf{25,}
236 (1973).

\bibitem{hu}W. Hu and M. White, arxiv:astro-ph/9702170 (1997).

\bibitem{kuijla}A. B. Kuijlaars, A. Martinez-Finkelshtein and R.
Orive, \textit{ETNA} \textbf{19,} 1 (2005).

\bibitem{Sharma}S. K. Sharma, P. R. Dhungel and U. Khanal, arXiv:
1307.5443v1 {[}astro-ph.CO{]}.

\bibitem{mathew}J. Mathews and R. L. Walker, \textit{Mathematical
Methods of Physics}, Benjamin, Menlo Park (1970).

\bibitem{grads}I. S. Gradshteyn and I. M. Ryzhik, \textit{Table of
Integral, Series and Products,} Academic, New York (1980).

\bibitem{Kolb-Turner}E. W. Kolb and M. S. Turner, \textit{The Early
Universe}, (Addison-Wesley, New York, 1990).

\bibitem{dhungel-khanal}P. R. Dhungel and U. Khanal, arXiv: arXiv:1109.6412v2
{[}astro-ph.CO{]}.

\bibitem{alisauskaus}S. Alisauskaus, \textit{J. Phys. A: Math. Gen.
}\textbf{35}, 7323 (2002).

\end{thebibliography}
\end{document}